\documentstyle[12pt]{article}
\begin{document}
\Large
\hfill\vbox{\hbox{DTP/94/24}
            \hbox{March 1994}}
\nopagebreak

\vspace{0.75cm}
\begin{center}
\LARGE
{\bf The Three-Jet Rate in \(e^{+}e^{-}\) Annihilation}
\vspace{0.6cm}
\Large

C.N.Lovett-Turner

\vspace{0.4cm}
\large
\begin{em}
Centre for Particle Theory, University of Durham\\
South Road, Durham, DH1 3LE, England
\end{em}

\vspace{1.7cm}

\end{center}
\normalsize
\vspace{0.45cm}
\begin{em}
Progress has been made on the calculation of \(R_{3}\), the three-jet
rate
in \(e^{+}e^{-}\) annihilation, in the \(k_\bot \) (Durham) scheme.
Using the
coherent branching formalism \cite{b,c,d}, an explicit
expression for \(R_{3}\) is calculated. In this, leading and
next-to-leading
large logarithms (LL and NLL) are resummed to all orders in QCD
perturbation
theory. In addition to exponentials an error function is involved.
\end{em}
\newpage

The process \(e^{+}e^{-} \) annihilation to hadronic jets has proved
itself amply as a powerful tool for the analysis of QCD. Yet, as it is
 always
jets which are observed rather than the QCD objects in which one is
interested,
it is necessary to have some mechanism for deciphering useful
information from
the observed hadronic final states. The \(k_\bot \) or Durham
algorithm \cite{a}
is just one of several jet clustering algorithms which are used to
define jets.
As the name suggests it is based on the transverse momenta of the
observed
hadrons and works as follows:
\begin{enumerate}
\item Define a resolution parameter \(y_{cut} \)
\item For every pair of hadrons \(h_{k}, h_{l} \) compute
\[y_{kl}= \frac{2(1-cos\theta_{kl})}{Q^{2}} min(E_{k}^{2},E_{l}^{2})
\]
\item If \(y_{ij} \) is the smallest value of \(y_{kl} \) computed
in 2.\ and
\(y_{kl}<y_{cut} \) then combine the momenta \(p_{i}, p_{j} \) into
a single jet
(`pseudoparticle') \(p_{ij} \) according to some recombination
prescription.
\item Repeat this procedure from step 2. until all pairs of
particles and/or
pseudoparticles have \(y_{kl}>y_{cut} \). Whatever objects now
remain are called
jets.
\end{enumerate}

This fulfills the conditions that the algorithm should be infrared and
 collinear
safe and that it should be subject to small hadronization corrections.
Work by Catani, Webber et al.\cite{a} has given implicit integral
expressions
for the jet rates \(R_{n}(y_{cut}) \) which will exactly sum leading
and
next-to-leading powers of ln$(1/y_{cut})$ to all orders of
perturbative QCD.
However, what is required for phenomenological applications is a set
of explicit
expressions. Numerical evaluation of the integral equations mentioned
above has
provided some useful information and has also acted as a valuable
check for our
results.

\indent The broad aim of recent work has been to find a general
expression for
the \mbox{$n$-jet} rate \(R_{n}(y_{cut}) \) and this is most easily
expressed in
terms of the generating function \(\phi(Q,Q_{0};u) \) where
\begin{equation}
Q_{0}^{2} = y_{cut}Q^{2}
\end{equation}
and {\em u} is a jet label. The $n$-jet rate is then
\begin{equation}
R_{n}(y_{cut}) = \frac{1}{n!}\left(\frac{\partial}{\partial
u}\right)^{n}\phi(Q,Q_{0};u)\biggr|_{u=0}
\end{equation}
 From the application of the coherent branching formalism \cite{b,c,d},
the generating function
obeys the following implicit coupled equations \cite{a}:
\begin{equation}
\phi(Q,Q_{0};u) = u^{2}\mbox{exp}\left(2\int_{Q_{0}}^{Q} dq
\Gamma_{q}(Q,q)[\phi_{g}(q,Q_{0};u) -1]\right)
\end{equation}
and
\begin{eqnarray}
\phi_{g}(Q,Q_{0};u)&=&u\mbox{exp}\biggr(\int_{Q_{0}}^{Q}dq\{
\Gamma_{g}(Q,q)[\phi_{g}(q,Q_{0};u)-1]-\Gamma_{f}(q)\}\biggr).
\nonumber\\
& &\biggr(1+u\int_{Q_{0}}^{Q}dq\Gamma_{f}(q)\mbox{exp}\biggr(
\int_{Q_{0}}^{q}dq'\{[
2\Gamma_{q}(q,q')-\Gamma_{g}(q,q')].\nonumber\\
& &[\phi_{g}(q',Q_{0};u)-1]+\Gamma_{f}(q')\}\biggr)\biggr)
\end{eqnarray}
Where the emission probabilities are defined as
\begin{eqnarray}\Gamma_{q}(Q,q) &=& \frac{2C_{F}}{\pi}
\frac{\alpha_{s}(q)}{q}
\left(\mbox{ln}\frac{Q}{q}-\frac{3}{4} \right)\\
\Gamma_{g}(Q,q) &=& \frac{2C_{A}}{\pi} \frac{\alpha_{s}(q)}{q}
\left(\mbox{ln}\frac{Q}{q}-\frac{11}{12} \right)\\
\Gamma_{f}(q) &=& \frac{N_{f}}{3\pi} \frac{\alpha_{s}(q)}{q}
\end{eqnarray}
For $R_{2}(y_{cut})$ one has a simple exponentiation and an
explicit expression
(8) for the resummed leading and next-to-leading logarithms
\cite{a,e} is
straightforward to obtain:
\begin{equation}
R_{2}=\mbox{exp}\left[\frac{C_{F}aL}{2}(3-L)-\beta_{0}
\frac{C_{F}a^{2}L^{3}}{12}
\right]
\end{equation}
where $L=$ ln$(1/y_{cut})$, $a=\alpha_{s}(Q)/\pi$. Applying
the formalism to
the case of \(R_{3}(y_{cut}) \) one finds that
\begin{equation}
R_{3}=2R_{2}\int_{Q_{0}}^{Q}dq\Gamma_{q}(Q,q)\mbox{exp}
\left[-\int_{Q_{0}}^{q}
dq'(\Gamma_{g}(q,q')+\Gamma_{f}(q'))\right]
\end{equation}
Substituting the emission probabilities into equation (9) -
and bearing in mind
the running of $\alpha_{s}$ - gives a virtually intractable
nested integral.
There are also no obvious means by which to isolate before
integrating terms
which will contribute to LL or NLL parts of $R_{3}$. Our approach
centred on
performing the calculation in two easily tractable stages.

Firstly the integral equation was evaluated with $\beta_{0}$ set
to zero to
give the terms in $R_{3}$ which are independent of $\beta_{0}$.
\begin{eqnarray}
R_{3}^{0}&=&4C_{F}aR_{2} \int_{Q_{0}}^{Q}\frac{dq}{q}\left(\mbox{ln}
\frac{Q}{q}
-\frac{3}{4}\right) \mbox{exp} \left[-2C_{A}a\int_{Q_{0}}^{q}dq'
\frac{1}{q'}
\mbox{ln}\frac{q}{q'}\right]\\
&=&-\frac{C_{F}aL}{2}\mbox{exp}\left[-\frac{C_{F}aL^{2}}{2}\right]
\Biggr\{
\frac{1}{2}\sqrt{\frac{\pi}{A}}\mbox{erf}(\sqrt{A})
(3(1-C_{F}aL^{2})-2L)
\nonumber\\& &+\frac{2}{C_{A}}(1-\mbox{exp}[-A])\left(
\frac{2}{aL}+3C_{F}\right)
\Biggr\}
\end{eqnarray}
where $A=C_{A}aL^{2}/4$ and erf$(x)$ is defined by
\begin{equation}
\mbox{erf}(x)=\frac{2}{\sqrt{\pi}}\int_{0}^{x}e^{-t^{2}}dt
\end{equation}
Note that we have used $\beta_{0}=(11C_{A}-2N_{f})/3$ to write
$\Gamma_{f}$ in
terms of $\beta_{0}$.

Secondly
\begin{eqnarray}
R_{3}^{'}&=&\frac{\partial R_{3}}{\partial\beta_{0}}
\Biggr|_{\beta_{0}=0}
\nonumber\\&=&\frac{\partial}{\partial\beta_{0}}\Biggr(2R_{2}
\int_{Q_{0}}^{Q}dq\frac{2C_{F}\alpha_{s}(q)}{\pi q}
\left(\mbox{ln}\frac{Q}{q}-\frac{3}{4}\right)\nonumber\\& &\mbox{exp}
\left[\int_{Q_{0}}^{q}dq'\frac{2C_{A}\alpha_{s}(q')}{\pi q'}\mbox{ln}
\frac{q}{q'}-\frac{\beta_{0}\alpha_{s}(q')}{2\pi q'}\right]\Biggr)
\Biggr|_{\beta_{0}=0}
\end{eqnarray}
with
\begin{equation}
\alpha_{s}(Q)=\frac{\alpha_{s}(\mu)}{1+\frac{\beta_{0}
\alpha_{s}(\mu)}{4\pi}
\mbox{ln}\frac{Q}{\mu}}
\end{equation}
was evaluated, thus giving the coefficients of the terms in
$R_{3}$ which are
proportional to $\beta_{0}$. This approach hinges on the
assumption that $R_{3}$
at NLL level is linear in $\beta_{0}$. To justify this
\begin{equation}
R_{3}^{''}=\frac{\partial^{2} R_{3}}{\partial\beta_{0}^{2}}
\Biggr|_{\beta_{0}=0}
\end{equation}
was evaluated and was found to contain only subleading terms
(that is terms
$\sim a^{n}L^{2n-p}$ with $p>1$), hence validating our method.

Any subleading terms which appeared as a result of this procedure
were discarded
since they would have necessarily been incomplete. Combining the
results of the
two stages according to
\begin{equation}
R_{3}=R_{3}^{0}+\beta_{0}R_{3}^{'}
\end{equation}
gives:
\begin{eqnarray}
R_{3}(y_{cut})&=&-\frac{C_{F}a L}{2}\mbox{exp}\Biggr[
-\frac{C_{F}a L^{2}}{2}
\Biggr]\nonumber\\
& &\Biggr\{ \frac{1}{2}\sqrt{\frac{\pi}{A}}\mbox{erf}
( \sqrt{A})\Biggr[3-2L+\frac{\beta_{0}
aL^{2}}{2}\Biggr(\frac{C_{F}a L^{2}}{3}-\frac{1}{2}\Biggr)
-3C_{F}a L^{2}+\frac{\beta_{0}}{2C_{A}}\Biggr]\nonumber\\
& &+\frac{1}{C_{A}}
\Biggr(\mbox{exp}[-A]
-1\Biggr)\Biggr[\frac{C_{F}\beta_{0}
aL^{2}}{3}-\frac{\beta_{0}}{6}-\frac{4}{a L}-6C_{F}\Biggr]-
\frac{\beta_{0}}
{2C_{A}}
\Biggr\}
\end{eqnarray}
This has been expressed in terms of error functions and exponentials
for
purposes of familiarity; but it should be noted that an equally
succinct
representation in terms of degenerate hypergeometric functions is
possible
and work on LL expressions for higher jet rates has shown that these
functions
appear to occur naturally.

One might speculate on the possibility of making a change of
renormalisation scale such that one could absorb any terms
proportional to
$\beta_{0}$. This can be done for $R_{2}$ where the $\beta_{0}$ term
in equation
(8) can be removed by replacing $\alpha_{s}(Q)$ by
$\alpha_{s}(y_{cut}^{1/3}Q)$. A closer look at equation (11),
however, reveals that $\beta_{0}$ enters $R_{3}$ through a combination
 of the
emission probabilities as well as through the renormalisation scale;
so it seems
likely that no scale can be chosen so as to remove all $\beta_{0}$
dependence.

In conclusion, we have found an explicit form for $R_{3}(y_{cut})$
which resums
leading and next-to-leading large logarithms, ln$(1/y_{cut})$, to all
 orders in
perturbative QCD. Phenomenological applications of equation (17) will
 be
discussed in forthcoming work.

\section*{Acknowledgments}
I would like to thank S.E.R.C. for a U.K. Studentship.

\end{document}